Tunneling Hamiltonian representation of false vacuum decay
I. comparison with Bogomol'nyi inequality

A. W. Beckwith
*Department of Physics and*
*Texas Center for Superconductivity and Advanced Materials*
*University of Houston*
*Houston, Texas 77204-5005, USA*

## ABSTRACT

The tunneling Hamiltonian has proven to be a useful method in many-body physics to treat particle tunneling between different states represented as wave functions. Here we present a generalization of the tunneling Hamiltonian to quantum field theory in which tunneling between states represented as wave functionals of a scalar quantum field $\phi$ is considered. We examine quantum decay of the false vacuum in the driven sine-Gordon system and show that it is consistent with the tunneling formalism derived here and matches up with the S-S' (S-S') separation obtained from the Bogomol'nyi inequality. This inequality permits construction of a Gaussian wave functional representation of S-S' nucleated states and is consistent with the false vacuum hypothesis.

CORRESPONDENCE:
A. W. Beckwith:      **projectbeckwith2@yahoo.com**

PACS numbers: 03.75.Lm, 11.27.+d, 71.45.Lr, 75.30.Fv , 85.25.Cp

# INTRODUCTION

In this paper, we use the vanishing of a topological charge Q to show how the Bogomol'nyi inequality can be used to simplify a Lagrangian potential energy term so that the potential energy is proportional to a quadratic $\phi^2$ scalar field contribution. In doing so, we work with a field theory featuring a Lorenz scalar singlet valued field in D+1 dimensional spacetime.

One should point out that such topological charges and inequalities, exist and hold respectively, for D+1 dimensional theories featuring scalar and singlet valued fields, only for D = 1. For D > 1, the Lorentz scalar fields must be D-plets! We use the D = 1 dimensional case for describing the dynamics of quasi one-dimensional metallic materials in our condensed matter example. Furthermore, we use the Bogomol'nyi inequality to obtain an appropriate wave functional that is then be put into a tunneling Hamiltonian. How we write this wave functional also makes extensive use of the quantum decay of the false vacuum hypothesis as well as the vanishing of topological charge mentioned above.

We then describe how the quantum decay of a false vacuum contributes to our problem. This is extremely important because Sidney Coleman used a least action principle for WKB-style modeling of tunneling, which we will use in part for forming the wave functionals in our new functional integration presentation of transport theory .

The quantum decay of the false vacuum[1] has been of broad scientific interest for over two decades. Several quantum tunneling approaches to this issue have been proposed. One [i] is to use functional integrals to compute the Euclidean action ("bounce") in imaginary time. This permits one to invert the potential and to modify what was previously a potential barrier separating the false and true vacuums into a potential well in Euclidean space and imaginary time. The decay of the false vacuum is a potent paradigm for describing decay of a metastable state to one of lower potential energy. In condensed matter, this decay of the false vacuum method has been used[2] to describe nucleation of cigar-shaped regions of true vacuum with soliton-like domain walls at the boundaries in a charge density wave. We use the Euclidian action so that we may invert the potential in order to use WKB semiclassical procedures for solving our problem. Another approach[3], using the Schwinger proper time method, has been applied by other researchers to calculate the rates of particle-antiparticle pair creation in an electric field[4] for the purpose of simplifying transport problems.

The tunneling Hamiltonian[5,6] involves matrix elements for modeling the transfer of particles between initial and final wave function states. Josephson[7] employed the tunneling Hamiltonian in his theory of phase-coherent tunneling of Cooper pairs through an insulating barrier. However, the tunneling Hamiltonian procedure has not previously been done in a way that allows field theoretic wave functionals to represent nucleating states from a vacuum state that is put into an evaluation of kinetic dynamics directly. The functional tunneling Hamiltonian is especially useful because it permits putting potential energy information into the wave functionals and analyzing the kinetics of the evolution between initial and final wavefunctional states with a minimum of algebraic manipulations when one considers systems of many weakly coupled fields. For example, Hawking et al.[8] point out that a universe can be nucleated by a cosmological instanton that is much larger than the Planck scale, provided there are sufficiently many matter fields. We should note that we can also use imaginary time to present the Hawking temperature of a black hole as a quantum effect[9], which is a good example of how imaginary time can be used to obtain helpful physical information. We use wave functionals with Euclidian actions representing how our evaluated states nucleate from 'nothingness' to a physical entity, which we can present in terms of a least action principle.

We then put these wavefunctionals into a tunneling Hamiltonian matrix calculation to show how to get the current density of objects being transformed from one initial physical state to a final state via a kinetic analysis of their behavior when tunneling through the barrier. This analysis is used to simplify transport problems for systems with weakly coupled fields. Our method simplifies what were excessively complicated solutions and fits in well with more abstractly presented treatments of this idea[10]. We also mathematically elaborate upon the S-S' domain wall paradigm[11,12], giving specific geometrical reasons why quasi one-dimensional systems are particularly amenable to this field theoretic treatment of tunneling Hamiltonians.

## II A. BASIC TECHNIQUES USED IN THIS PAPER

In this study, we apply the domain wall physics of S-S' pairs to obtain a quadratic scalar valued potential for transport physics problems involving weakly coupled scalar fields. We found that it was necessary in the beginning to write up the energy of a soliton kink and then to apply the Bogomol'nyi inequality to obtain a greatest lower bound to an energy of the kink expression integrated over spatially (to obtain a 'mass' of this 'kink'), which is a topological charge. After this energy/mass representation of the soliton kink is modified by the Bogomol'nyi inequalitiy, we can use the bound on our modified potential to simplify a Euclidian least action integral by changing the coefficient outside the action integral from a complex number to a real minus number by means of the imaginary time procedure. Incidentally, this imaginary time designation was how Sidney Coleman inverted his false vacuum potential. This potential inversion was done to permit a semiclassical WKB analysis.

For a S-S' pair, the topological charge Q (so designated due to domain walls) vanishes. If we use Euclidian imaginary time, the least action integral of our wave functional will be changed from Eq. (1a) below to Eq. (1b) by using $time \to i \cdot (time)$.

$$\int D\phi \cdot \exp\left( (i/\hbar) \cdot \int d^d x \cdot \left[ \frac{1}{2} \cdot (\partial \phi)^2 - V(\phi) \right] \right) \to \qquad (1a)$$

transforms to

$$\int D\phi \cdot \exp\left( (-1/\hbar) \cdot \int d^d_E x \cdot \left[ \frac{1}{2} \cdot (\partial \phi)^2 + V(\phi) \right] \right) \qquad (1b)$$

We should note that Eq. (1b) has an energy expression of the form

$$\varepsilon(\phi) \equiv \int d^d_E x \cdot \left[ \frac{1}{2} \cdot (\partial \phi)^2 + V(\phi) \right] \qquad (2a)$$

Eq. (2a) has a potential term that we can write as

$$V(\phi) \equiv C_0 \cdot (\phi - \phi_0)^2 + C_1 \cdot (\phi - \phi_0)^4 + H.O.T. \qquad (2b)$$

Furthermore, even after we invert our potentials, we can simplify our expression for the potential by procedures that eliminate the scalar potential terms higher than $\phi^2$ by considering the energy per unit length of a soliton kink. This is given in A. Zees book as being, after rescaling to different constants,

$$\tilde{\varepsilon}(x) = \frac{1}{2} \cdot \left( \frac{d \cdot \phi}{d \cdot x} \right)^2 + \frac{\lambda}{4} \cdot (\phi^2 - \varphi)^2 \qquad (3)$$

with a mass of the kink or antikink of this given by

$$M \equiv \int dx \cdot \tilde{\varepsilon}(x) \qquad (3a)$$

to be bounded below, namely, by use of the Bogomol'nyi inequality

$$M \geq \int dx \cdot \sqrt{\frac{\lambda}{2}} \cdot \left|\left(\frac{d \cdot \phi}{d \cdot x}\right) \cdot (\phi^2 - \varphi^2)\right| \geq \left|\frac{4}{3 \cdot \sqrt{2}} \cdot \mu \cdot \left(\frac{\mu^2}{\lambda}\right) \cdot Q\right| \quad (4)$$

where Q is a topological charge of the domain wall problem. We define conditions for forming a wave functional via the Bogomol'nyi inequality and the vanishing of the topological charge Q, as given by Eq. (5):

$$\Psi = |0>^0 \equiv c \cdot \exp(-\alpha \cdot \int dx^{(D\equiv 1)} [\phi - \phi_C]^2) \quad (5)$$

We presuppose, when we obtain Eq. (5), a power series expanasion of the Euclidian Lagrangian, $L_E$ about $\phi_C$. The first term of this expansion,

$$L_E |_{\phi=\phi_O} = \frac{1}{2} \cdot (\vec{\nabla}\phi)^2 |_{\phi=\phi_0} \equiv \varepsilon(\phi)|_{\phi\equiv\phi_0} \quad (6)$$

is a comparatively small quantity that we may ignore most of the time. Furthermore, we simplify working with the least action integral by assuming an almost instantaneous nucleation of the S-S' pair. We may then write, starting with a Lagrangian density $\zeta$,

$$\int d\tau \cdot dx \cdot \zeta \to t_P \cdot \int dx \cdot \zeta \to value \equiv t_P \cdot \int dx \cdot L \quad (7)$$

Quantity $t_P$ in equation 7 is scaled to unity. Eq. (7) allows us to write our wave functional as a one-dimensional integrand.

Introducing domain wall physics via Eq. (6) and Eq. (7) allows us to use Sidney Coleman's 1977 introduction of a least action integral interpretation of WKB tunneling as the starting point to our analysis. This permits us to write our wave functional as proportional to

$$\psi \propto c \cdot \exp(-\tilde{\beta} \cdot \int L \, d\tau) \quad (8)$$

The wave functional so obtained is then put into a treatment of the tunneling Hamiltonian that we derive later. Due to the presence of S-S' pairs, the tunneling Hamiltonian matrix element scales as a super current proportional to the modulus of the tunneling Hamiltonian. In this case, the super current goes linearly with the effective matrix element for transferring a single boson. This is similar to how Feynman's derivation[17] of the Josephson current-phase relationship. Our analysis of the tunneling Hamiltonian matrix element via our modified wave functionals permits a current density expression to be proportional to the modulus of the tunneling Hamiltonian, |*T*|.

The conclusion of our derivation assumes our matrix element *T* contains the kinetic dynamics of our transport problem, while the wave functionals use the problem's assumed potential energy barrier. Our coefficient α outside the integrand of the wave functional is shown to use the fate of the false vacuum hypothesis.

## II b. What is the false vacuum hypothesis, and why it is so important to this problem?

First, envision a "true" vacuum, which is absolutely nothing—no energy, no matter. This vacuum, this nothing, however, is actually a whole lot of something. Due to the quirks of quantum mechanics, a vacuum is really a teeming froth of particles and antiparticles popping into existence and then annihilating each other. In a true vacuum, the energy states of these so-called virtual particles add up to zero. So, while a true vacuum may indeed be something other than nothing, it still does not have any energy in it. In a false vacuum, however, the combination of energy states has a nonzero value. So a vacuum, an area of the universe at its lowest energy state because it is completely devoid of any matter or radiation, could still have some energy associated with it. The value of this energy is called the cosmological constant.

The space inside a bubble of false vacuum is curved, or warped, and a small amount of energy is stored in that curvature, like the potential energy of a strung bow. This ostensible violation of energy conservation is allowed by the Heisenberg uncertainty principle for sufficiently small time intervals.

In any case, the false vacuum hypothesis can be visualized, as seen in Figure 1, as being a local potential minimum that is higher than the final minimized equilibrium state of our physical system. Cosmologists have been referring to this false vacuum paradigm for decades, since Sidney Coleman introduced the concept in 1977. Afterwards, Kazumi Maki was one of the first condensed matter physicists to apply the false vacuum hypothesis to help determine appropriate conditions for tunneling through a potential barrier for density wave systems. We are using the false vacuum hypothesis for giving values of the alpha coefficient which is placed in front of the integrand of our wave functional used in our tunneling Hamiltonian. This alpha is inversely proportional to both the difference in local minimum energy levels seen in Figure 1, upper right hand side, and to the distance between a S-S' pair found in a pinning gap in charge density wave transport through quasi one-dimensional materials.

## II c. Why use functional integration for this transport problem?

In physics, functional integration is integration over certain infinite-dimensional spaces. In applications to physics, *functional integration* refers to integration over spaces of paths or, more generally, over field configurations. We use this implicitly in our tunneling Hamiltonian construction in our variation over phase space $\phi$. Furthermore, we make extensive use of functional derivatives. Note that there are two ways to define functions, *implicitly* and *explicitly*. Most of the equations we are familiar with are explicit equations, such as y = 2x – 3, so that we can write y = f(x), where f(x) = 2x – 3. But the equation 2x – y = 3 describes the same function. This second equation is an implicit definition of y as a function of x. As there is no real distinction between the appearance of x or y in the second form, this equation is also an implicit definition of x as a function of y So, functional differentiation is, in many respects, similar to the idea of implicit differentiation in calculus, especially with respect to the examples given above. However, functional derivatives also often have one striking difference from ordinary calculus, in that we work with what is a delta function localization, where the $\delta(x-y)$ is infinite if we set x = y, whereas it is equal to zero elsewhere. We can represent this implicit dependence of functional differentiation by the following:

$$\frac{\delta\ f(x)}{\delta \cdot f(y)} \equiv \delta(x-y) \tag{9}$$

This has significant implications with respect to the physics of the tunneling Hamiltonian. Specifically, in our tunneling Hamiltonian we do an integration by parts, turning the delta function into a step function, which implies that the value of the phase $\phi_0$ will be between zero and $2 \cdot \pi$, as seen in the main diagram of Figure 1. Our functional derivatives make it easier to work with general basis functions $\phi(x)$, which we use in our transport problem. This is due to our write up of

$$\frac{\delta \cdot V(x(t'))}{\delta\ x(t)} \equiv \lim_{\varepsilon \to 0} \frac{V(x(t') + \varepsilon \cdot \delta(t'-t)) - V(x(t'))}{\varepsilon} \tag{10}$$

as a generalized functional derivative used in the tunneling Hamiltonian calculations.

## II d. Driven sine Gordon equation used in this presentation

We are working with physical systems with a dominant potential of the form

$$V(\phi) \equiv (1 - \cos\phi) \tag{11}$$

with a small perturbing potential added. This gives the physical system a slightly tilted double well potential characteristic, with an external field pushing solitons (antisolitons) through the dominant potential given in Eq. (11) above. Such is the case when we work with the washboard potential model of soliton transport in classical systems, with our small quantum perturbing force acting on soliton (antisoliton) pairs. We usually make a $\phi^4$ round off of quartic contributions to the potential we are working with. This permits use of the prior approximations, e.g., the Bogomol'nyi inequality, in our formation of Gaussian wave functionals used to permit nucleated states as given by the main representation, to the left, in Figure 1.

## II e. Plan of this paper

In our paper, we will initially refer to the classical, time-dependent derivation of tunneling Hamiltonians from the standpoint of typical Fermi Golden Rule calculations. Section III outlines traditional approaches and includes a discussion of how coherent tunneling of bosons leads to a critical current proportional to the modulus of a tunneling Hamiltonian matrix element, |T|. However, for reasons we bring up in our discussion, we use the Bogomol'nyi inequality[13] (Section V) with wavefunctionals mimicing a Gaussian in a revamped functional form of Tekemans representation of a field theoretic matrix element (Section VI). The difference between a phi to the fourth power potential contribution and the mass of a kink (here we treat a S-S' pair as a kink-antikink in spacetime) leads us to give a mathematical statement of the effects of quantum entanglement for S-S' nucleation applications (Appendix I), which reinforces the physical argument given in Section VII. Section VII has a proof of how the Bogomol'nyi inequality[13] ties in with the false vacuum model for S-S' pairs[11]. Section VII gives us wavefunctionals, which in momentum space leads (as shown in our next paper's explicit CDW [charge density wave] example) to a precise data matchup with experimental measurements for an *I-E* curve (we are referring to a 1985 experiment with an applied electric field generating a current representing CDW transport). Section VIII is our conclusion.

### III. TUNNELING HAMILTONIAN IN QUANTUM MECHANICS

The underpinnings of the tunneling Hamiltonian method began with Oppenheimer's study[14] of field-induced ionization of the hydrogen atom. What we now call the tunneling Hamiltonian was proposed by Bardeen[6] to treat tunneling of many electrons through a barrier separating two electrodes. The classical theory of tunneling Hamiltonians has been frequently applied to condensed matter issues, and the result is usually for a constant value of matrix elements Tekman[15] has extended the tunneling Hamiltonian method to encompass more complicated geometries, such as the tips used in scanning tunneling microscopy. In this case, the transmission rate can be calculated using first-order time-dependent perturbation theory[16]. The result, for incoherent quasiparticle tunneling, This usually is for

$$P = \frac{2\pi}{\hbar} \sum_{k,q} |T_{kq}|^2 \delta(E_k - E_q) \qquad (12)$$

involving

$$T_{kq} = -\frac{\hbar^2}{2m} \int_{S_0} \left[\psi_k^* \nabla \psi_q - \psi_q \nabla \psi_k^*\right] \cdot d\mathbf{S} \tag{13}$$

In addition, we should note that the tunneling Hamiltonian matrix elements used the second quantized formalism by Cohen, Falicov, and Phillips[5] and later used by Josephson[7] in his landmark paper on coherent tunneling of paired electrons between two superconductors. For our updated formalism, we had $\psi_k$ and $\psi_q$ ($E_k$ and $E_q$) represent wavefunctions (eigenenergies) for states in infinitely separated left and right electrodes, respectively. The electrodes are imagined as being brought into close proximity, and the integration is carried out over an arbitrary surface $S_0$ that lies inside the classically forbidden region, where the wave functions decay (eg. exponentially) through the potential energy barrier. For the case of 1-D tunneling through a rectangular barrier ranging from $x_a$ to $x_b$ along the x-direction, Eq. (2) becomes:[6]

$$\begin{aligned} T_{kq} &= -\frac{\hbar^2}{2m} \int_{x_a}^{x_b} dx \left[\psi_k^* \frac{\partial \psi_q}{\partial x} - \psi_q \frac{\partial \psi_k^*}{\partial x}\right] \delta(x-x_0) \\ &= \frac{\hbar^2}{2m} \int_{x_a}^{x_b} dx \left[\psi_k^* \frac{\partial^2 \psi_q}{\partial x^2} - \psi_q \frac{\partial^2 \psi_k^*}{\partial x^2}\right] \theta(x-x_0), \end{aligned} \tag{14}$$

where $\theta(x - x_0)$ is the unit step function and $x_0$ is an arbitrary point within the barrier. In the TH formulation, tunneling is represented by the transfer of particles from states $k$ on the left side to states $q$ on the right side of the barrier (or vice-versa). This leads, using second quantized formalism, to the following additional term, known as the tunneling Hamiltonian,[5] in the total system Hamiltonian:

$$H_T = \sum_{k,q} T_{kq} \left[c_q^+ c_k\right] + h.c. \tag{15}$$

where $c_q^+$ and $c_k$ usually represent fermion creation and annihilation operators. When the matrix elements $T_{kq}$ are small, the current through the barrier is calculated using linear response theory, and is found to be proportional to $|T|^2$ for quasiparticle tunneling, as suggested by equation (13). We should note that this may be used to describe coherent, Josephson-like tunneling of either Cooper pairs of electrons or boson-like particles, such as superfluid $^4$He atoms. In this case, the supercurrent goes linearly with the effective matrix element for transferring a pair of electrons or transferring a single boson, as shown rather elegantly in Feynman's derivation[17] of the Josephson current-phase relation. This means a current density proportional to $|T|$ rather than $|T|^2$ since tunneling, in this case, would involve coherent transfer of individual bosons (first-order) rather than pairs of fermions. metals As a result, the Josephson critical current for coherent pair tunneling is proportional to $|T|^2$ rather than $|T|^4$ (which would have been the case for incoherent tunneling since pair tunneling is a second-order process)[7] In either case, we will be able to use the Bogomol' yi inequality [13] in order to isolate a Gaussian contribution to the wave functional states used in our field theoretic tunneling Hamiltonian. In our investigation we found that the instanton approach for nucleation of *objects* from '*nothing*' into a general space dS (e.g. as Lemos[18] called dS the de Sitter metric space) as compared to the cumbersome 2$^{nd}$ quantization procedure above allows us to present the physics of this problem without involving the complicated structures of 2$^{nd}$ quantization which may not be appropriate _for , as an example, S-S$^{'}$pairs traversing a pinning gap in quasi one dimensional metals.

## IV. MOTIVATION FOR USING WAVE FUNCTIONAL REPRESENTATION

It is straightforward to form a representation of wave functionals which represent creation of a particular event within an embedding space. Diaz and Lemos[18] use this technique, as an example of the exponential of a Euclidian action applied to show how black holes nucleate from 'nothing'. This was done in the context of de Sitter space, and in addition, Diaz and Lemos[18] used a similar calculation with respect to nucleating a de Sitter space from 'nothing'. The ratio of the modulus of these two wave functionals is used to calculate the 'probability' of Black holes being nucleated within a de Sitter space which is the general embedding space of the 'universe'. This trick was also used by Kazumi Maki[19] in 1977-8 when he looked at a field theoretic integration of condensates of S-S' pairs in the context of boundary energy of a two dimensional bubble of 'space-time'. This two dimensional bubble action value was _minus a contribution to the action due to volume energy of the same 2 dimensional bubble of 'space- time,' Maki[19] came up with a probability expression for S-S' pair production not materially different from what Diaz and Lemos[18] used for their black hole nucleation problem. Further arguments by Maki[19] in 1978 derived the probability for S-S' pair production as very similar to a standard Zenier expression which qualitatively is similar to what was used by Lin[20] for his derivation of electron-positron pair creation in D+1 dimensions, where he recovered for pure electric fields the old Schwinger results.[3] .The problem is that these expressions in the case of S-S' pair production lead to qualitative graphs which do not fit experimental data, as seen in 1985.[21] .This was the starting point of our inquiry which we generalized to this article, where we corrected for incomplete earlier Zenier current derivations.[22] .The I-E curve in question will be in our next article of this series.

# V. APPLICATION OF BOGOMOL'NYI INEQUALITY TO WAVEFUNCTIONALS

In order to obtain a Gaussian approximation to a wave functional treatment of the dynamics of a potential with weakly coupled scalar fields, we start with a field treatment of the least action wavefunctional we shall set as Eq. (9) above, that is, $\psi \propto c \cdot \exp(-\beta \cdot \int L \, d\tau)$, with an extremal 'action type behavior for the integral being subject to 2$^{nd}$ order variations in the potential. We represent the Lagrangian in terms of a power series expansion of the form, assuming that the first derivative of L vanishes

$$L = L|_{\phi=\phi_O} + \frac{1}{2} \cdot (\phi - \phi_0)^a (\phi - \phi_0)^b \cdot \frac{\partial^2 \cdot L}{\partial \cdot \phi^a \cdot \partial \cdot \phi^b} + \left(\text{neligible H.O.T.}\right) \quad (16)$$

We should be aware that for a wick rotation, when $t = -i \cdot \tau_E$ that for $d$ dimensions $d^d x = -i \cdot d_E^d x$ with $d_E^d x = d\tau_E \cdot d^{d-1} x$ [13,23]. Then $(\partial \phi)^2 = (\partial \phi / \partial t)^2 - (\vec{\nabla}\phi)^2$ becomes $(\partial \phi)^2 = (\partial \phi / \partial \tau_E)^2 + (\vec{\nabla}\phi)^2$ while we transform the problem to one of the form given by equations (1a) and (1b) above, which involves using a static energy functional of the field $\phi(x)$, namely by using the Euclidian energy $\varepsilon(\phi) \equiv \int d_E^d x \cdot \left[\frac{1}{2} \cdot (\partial \phi)^2 + V(\phi)\right]$ as seen in equation (2a) above. We may link this to quantum statistical physics via noting that Euclidian quantum field theory in ( D+1) dimensional space time, when $0 \leq \tau < \bar{\beta}$ is ~ Quantum statistical mechanics in D dimensional space. We will be picking a dominant quadratic contribution of $\tilde{C}_1 \cdot (\phi - \phi_0)^2$ from an action integral. Here, $\tilde{C}_1$ is a contribution we shall derive from an argument we present below, for potentials of the form, due to setting D=1, as we have already seen in Eq. (2b) above, i.e. $V(\phi) \equiv C_0 \cdot (\phi - \phi_0)^2 + C_1 \cdot (\phi - \phi_0)^4 + H.O.T.$. Now, if we proceed, via the 'false vacuum' methodology of Sidney Coleman , this then becomes[24] after we take a one dimensional D=1 version of a multi dimensional Taylor series expansion of the potential

$$L_E \approx L_E |_{\phi=\phi_O} + \frac{1}{2} \cdot (\phi-\phi_0)^2 \frac{\partial^2 \cdot V_E}{\partial \cdot \phi^2}|_{\phi=\phi_0} + \frac{1}{3!}(\phi-\phi_0)^3 \cdot \frac{\partial^3 \cdot V_E}{\partial \cdot \phi^3}|_{\phi=\phi_0} +$$

$$\frac{1}{4!}(\phi-\phi_0)^4 \cdot \frac{\partial^4 \cdot V_E}{\partial \cdot \phi^4}|_{\phi=\phi_0} \qquad (17)$$

We have a good approximation to simplifying equation (17) by using Eq. (8) above, i.e.,

$L_E |_{\phi=\phi_O} = \frac{1}{2} \cdot (\vec{\nabla}\phi)^2 |_{\phi=\phi_0} \equiv \varepsilon(\phi)|_{\phi\equiv\phi_0}$ .Here, we are making the experimentally justifiable assumption of denoting. $(\partial_E \phi)^2 \propto const. \to 0$ and assume we deal with the spatial contribution as varying but bounded .So we isolate the quadratic contribution to our wave functional we use the Bogomol'nyi inequality[13] while we have an energy per unit length of , in one dimension for a topological ' soliton ' kink which is given in Eq. (3) above, namely by use of $\tilde{\varepsilon}(x) = \frac{1}{2} \cdot \left(\frac{d \cdot \phi}{d \cdot x}\right)^2 + \frac{\lambda}{4} \cdot (\phi^2 - \varphi)^2$ for a soliton kink . We also use a conserved current quantity of [13,25]

$$J^\mu = \frac{1}{2 \cdot \varphi} \cdot \varepsilon^{\mu\nu} \cdot \partial_\nu \cdot \phi \qquad (18)$$

with a *topological* charge of [13]

$$Q \equiv \int_{-\infty}^{+\infty} dx \cdot J^0(x) = \frac{1}{2 \cdot \varphi}[\phi \cdot (\infty) - \phi \cdot (-\infty)] \qquad (19)$$

Note here that the denominator $\varphi$ is not the same as $\phi(x)$ ! In A. Zees book, the $\varphi$ term is due to his setting of two minimum positions for $\phi$ for a double well potential. He calls these v in his equation 1 of page 278 of his tome. We can use much of this same construction because our overall potential still will look like a double well potential but being tilted by application of an external field. Furthermore, we have that topological charge as defined above is similar to results where the lattice topological charge associated with Ginsparg-Wilson fermions exhibit generic topological stability over quantum ensemble of configurations contributing to the QCD path integral. Moreover, the underlying chiral symmetry leads to the suppression of ultraviolet noise in the associated topological charge densities. Using these tools it was recently demonstrated that: (a) there is a well-defined space-time structure (order) in topological charge density for typical configurations contributing to QCD path integral; (b) this fundamental structure is low-dimensional, exhibiting sign-coherent behavior on subsets of dimension less than four and not less than one; (c) the structure has a long-range global character (spreading over maximal space-time distances) and is built around the locally one-dimensional network of strong fields (skeleton). This is currently how they fit into many physical situations in field theory. Now, following the conventions given by Zee in "Quantum Field Theory in a Nutshell " we have that if we have meson type behavior for the *field* $\phi(x)$, this charge will vanish, and it is useful to note that if we look at the mass of a kink via a scaling $\mu \propto \sqrt{\lambda \cdot \phi_0^2}$ with *M* defined by being the same as the energy of a soliton kink as given in equation (3) with a subsequent mass given in equation (3a) above that then we have via using $a^2 + b^2 \geq 2 \cdot |a \cdot b|$ an inequality of the form given by Eq. (4) above , with

$$M \geq \int dx \cdot \sqrt{\frac{\lambda}{2}} \cdot \left|\left(\frac{d \cdot \phi}{d \cdot x}\right) \cdot (\phi^2 - \varphi^2)\right| \geq \left|\frac{4}{3 \cdot \sqrt{2}} \cdot \mu \cdot \left(\frac{\mu^2}{\lambda}\right) \cdot Q\right| \, , \quad \text{so that}$$

$$M \geq |Q| \tag{20}$$

with *mass* M in terms of units of $\frac{4}{3 \cdot \sqrt{2}} \cdot \mu \cdot \left(\frac{\mu^2}{\lambda}\right)$ If we note that we have, in one dimension, $(\phi - \phi_0)^4 = (\phi^2 - \phi_0^2)^2 - 4 \cdot \phi \cdot \phi_0 \cdot (\phi - \phi_0)^2$ we physically use our topological current as a vanishing quantity from the kinetic term and the fourth order term in an expansion of the potential about $\phi = \phi_0$. The Bogomol'nyi inequality thereby permits us to cancel non quadradic contributions to equation 8. We can say that this is the one spatial dimension plus one spatial dimension version of the follow situation. If we are considering a Hamiltonian system for a sine Gordon style potential in several dimensions of the form [26]

$$H_O = \int_x \left[\frac{1}{2} \cdot \Pi_x^2 + \frac{1}{2} \cdot (\partial_x \phi_x)^2 + \frac{1}{2} \cdot \mu^2 \cdot (\phi_x - \varphi)^2 - \frac{1}{2} \cdot I_0(\mu)\right] \tag{21}$$

we may obtain a 'ground state' wave functional of the form[26]

$$|0>^o = N \cdot \exp\left\{-\int_{x,y} (\phi_x - \varphi) \cdot f_{xy} \cdot (\phi_y - \varphi)\right\} \tag{22}$$

where we have due to higher order terms in a perturbing potential $H_1$

$$\frac{\partial^2 \cdot V_E}{\partial \cdot \phi^a \cdot \partial \cdot \phi^b} \propto f_{xy} \tag{23}$$

as this becomes equivalent to a 'coupling term' between the different 'branches' of this physical system. We restricted our analysis to quasi- one-dimensional cases via the following argument: a de facto 1+1 dimensional problem in transport physics ( via assuming an infinitesmal jump in a time $t_P$ = unit of Planck time length) to being one which is quasi one dimensional by making the following substitution, namely looking at the lagrangian density $\varsigma$ to having a time independent behavior denoted by a sudden pop up of a S-S' pair via the substitution of the nucleation 'pop up' time by what we did in Eq. (7) above, i.e. $\int d\tau \cdot dx \cdot \varsigma \to t_P \cdot \int dx \cdot L$, where we have that $t_P$ here is the Planck's time interval. Then afterwards, we shall use the substitution of $\hbar \equiv c \equiv 1$ so we can write

$$\int d\tau \cdot dx \cdot \varsigma \to t_P \cdot \int dx \cdot L \equiv G \cdot \int dx \cdot L \tag{24}$$

where

$$M_P \equiv \frac{1}{\sqrt{G}} \equiv 1.22 \times 10^{19} \, GeV = .231 \times 10^{20} \cdot m_e \tag{25}$$

such that

$$m_e \equiv 4.338 \times 10^{-20} \cdot M_P \to 4.338 \times 10^{-20} \tag{26}$$

So, if we make the substitution that $M_P \equiv 1 \Rightarrow G \equiv 1$ as a normalization procedure, we have we can indeed make the least action simplification alluded to in Eq. (7) and Eq. (24). This allowed us to use in our generalized nucleation problem the following wave functional

$$\psi \propto c \cdot \exp\left(-\beta \cdot \int L \, dx\right) \tag{27}$$

Assuming this is true, then we would be able to observe a 'ground state' via Eq. (5)

where we defined

$$\phi_C \equiv {}^o\!<0|\phi_x|0>^o \tag{28}$$

where the $|0>$ is for a ground ( or a vacuum ) state and Eq. (5) and Eq. (28) is for a standard model physics trajectory with field evaluated at $x \equiv (x^1,....,x^D)$ being a position in $D$ dimensional space then being set to $D = 1$. This means a wavefunctional with information from a inverted potential as part of a transport problem of weakly coupled systems along the lines suggested by Tekeman.[15],[27] We found our weakly coupled systems eliminated the cross terms in our derivation of a functional integral and for $D = 1$, can write more generally the initial configuration of the form[28] :

$$\Psi_i[\phi(\mathbf{x})]\big|_{\phi \equiv \phi_{Ci}} = c_i \cdot \exp\left\{-\alpha \int d\mathbf{x} \left[\phi_{ci}(\mathbf{x}) - \phi_0\right]^2\right\}, \qquad (29a)$$

or

$$\Psi_i[\phi(\mathbf{x})] = c \cdot \exp\left\{-\alpha' \int d\mathbf{x}\, L_E(\mathbf{x})\right\}$$
$$= c \cdot \exp\{-\alpha' S_E\}. \qquad (29\,b)$$

where we are looking at situations as shown in Figure 1. We are assuming that our

[ put **Figure 1** about here ]

initial state is similar to Colemans[1] false vacuum 'bounce' representation.[1] Next one includes an additional potential well representing the true vacuum, which is now separated by a potential barrier from the false vacuum, as shown in the upper right hand inset to Figure 1. The final state can be approximated as a modified Gaussian centered about a final field configuration of $\phi_{Cf}(x)$ that includes a bubble in which $\phi_0$ has tunneled through the barrier into the true vacuum state, creating one or more soliton domain walls at the boundary between true and false vacuums. Then for the final state immediately after tunneling,[13, 28] we write :

$$\Psi_f[\phi(\mathbf{x})]\Big|_{\phi \equiv \phi_{Cf}} = c_f \cdot \exp\left\{-\int d\mathbf{x}\, \beta(\mathbf{x})[\phi_{Cf}(\mathbf{x}) - \phi_0(\mathbf{x})]^2\right\}, \quad (29c)$$

where $\beta(\mathbf{x})$ may, in general, be complex. We in our CDW derivations kept $\beta(\mathbf{x})$ real valued and actually equal to $\alpha$ of equation 29a, but that is heavily dependent upon the basis $\phi$ we pick for our problem. In between the initial and final configurations, lies a family of curves $\phi_0(x)$ inside the tunnel barrier. In order to do this, we will work with wave functionals which obey the extremal condition of ( where $\phi_{ci,cf}(x)$ is the initial, final state equilibrium configuration of phase ).

$$\frac{\delta}{\delta\phi(x)}\left(\int L_{i,f}\, d\tau\right)\Bigg|_{\phi_0 \equiv \phi_{Ci,Cf}} \equiv 0 \qquad (30)$$

## VI. WAVE FUNCTIONAL REPRESENTATION OF TUNNELING HAMILTONIAN

The objective here is to transform Eq. (13) into an expression representing a scalar field $\phi(x)$. It is still possible considering a tunneling Hamiltonian matrix element $T_{if}$ that connects an initial state $|\psi_i\rangle \equiv \Psi_i[\phi]$ (before tunneling) with a final state $|\psi_f\rangle \equiv \Psi_f[\phi]$. The integration is changed, subsequently, to be over the 'area' of a tunneling barrier, $S_0$, which if we make a transformation based upon changing to a 'functional basis'[29] leads to

$$(T_{IF})_{functional}\Big|_{\Psi(real)} = \frac{-\hbar^2}{2\mu}\int\left(\Psi_{initial}^* \frac{\delta\Psi_{final}^*}{\delta\phi(x)} - \Psi_{final} \frac{\delta\Psi_{initial}^*}{\delta\phi(x)}\right)\delta(\phi(x) - \phi_0(x))\wp\,\phi(x) \quad (31)$$

Here, $dS \approx n \cdot dx$ where $n$ is the 'height' of the barrier between two 'half' regions. Also,

$$dx \to \frac{\delta x}{\delta \phi} \cdot \wp \phi(x) \tag{32}$$

and

$$\nabla \Psi_{i,f} \approx \frac{\partial}{\partial x} \Psi_{i,f} \to \left(\frac{\delta x}{\delta \phi}\right)^{-1} \frac{\delta \Psi_{i,f}}{\delta \phi}\bigg|_{\phi=\phi_0} \tag{33}$$

where $\left(\frac{\delta x}{\delta \phi}\right)^{-1}\bigg|_{\phi=\phi_0} \cdot \frac{\delta x}{\delta \phi} \equiv \frac{\delta \phi_0}{\delta \phi} \approx \delta(\phi - \phi_0)$ leads to [29], after $\psi_0 \to \Psi_{initial}$ and

$\psi_{mn} \to \Psi_{final}$ are put into Eq. (2) above.. We are actually working with a momentum '

representation of :

$$\pi(\mathbf{x})|\psi(t)\rangle \to -i\hbar \frac{\delta}{\delta \phi(\mathbf{x})} \Psi[\phi, t] \tag{34}$$

Requiring that the effective action $\Gamma = \int dt \langle \psi(t)|i\hbar \partial_t - H|\psi(t)\rangle$ be stationary against

arbitrary variations of $|\psi(t)\rangle \equiv \Psi[\phi, t]$ leads to the functional Schrödinger equation

$$i\hbar \frac{\partial \Psi[\phi, t]}{\partial t} = H\Psi$$

$$= \int_x \left[-\frac{\hbar^2}{2 \cdot m} \frac{\delta^2}{\delta \phi^2(\mathbf{x})} + \frac{1}{2}(\nabla \phi)^2 + V(\phi)\right] \Psi[\phi, t]. \tag{35}:$$

which is then used in our tunneling Hamiltonian in a change of variables leading up to

Eq. (31). We also found it useful to make an additional revision of Eq. (13) via

arguments similar to equation (14)

$$T_{if} \cong \frac{\hbar^2}{2\mu} \int \left(\Psi^*_{initial} \frac{\delta^2 \Psi_{final}}{\delta \phi(x)_2} - \Psi_{final} \frac{\delta^2 \Psi^*_{initial}}{\delta \phi(x)_2}\right) \vartheta(\phi(x) - \phi_0(x)) \wp \phi(x) \tag{36}$$

via use of $\vartheta(\phi(x)-\phi_0(x))$ as a step function. Before proceeding, we should note that the 'step function' $\vartheta(\phi(x)-\phi_0(x))$ tells us that the matrix element presented above is with respect to weakly coupled fields with the *trajectory* $\phi_0(x)$ inside a potential 'barrier' region and that we are interpreting $\psi_{initial}$ and $\psi_{final}$ as wave functional variations of quantum states about an initial and/or final configuration $\phi_{ci,cf}(x)$ state. For S-S' pairs this will be when $\phi_{ci}(x)|_{initial}$ is a nearly 'flat' state indicating pre nucleation values of the S-S' pair, whereas $\phi_{cf}(x)|_{final}$ is with regards to a fully formed S-S' pair. We can observe this in our right hand insert to Figure 1, which has $\phi_{ci}(x)|_{initial} \cong \phi_F$ and $\phi_{cf}(x)|_{final} \cong \phi_T$ where we will have $\phi_F$ as a false vacuum region and $\phi_T$ as a true vacuum region and where we will use the difference in energies between these two positions as being inversely proportional to the distance between a S-S' pair. Here we use $\phi(x)$ in whatever base we find convenient for this problem. In our applications, we used a DFT (discrete fourier transform) for $\phi(x)$ which allowed us to analyze the CDW problem in momentum space as having a gaussian wavefunctional form of presentation of a S-S', which is akin to what was done by Hermann G. Kümmel [28] who in 1998 used this concept that the soliton is "almost stable" up to the coupling parameter *beta* to form a solution to a sine – Gordon equation.

# VII. USE OF THE BOGOMIL'NYI INEQUALITY TO DECRIBE S-S' NUCLEATION

Let us begin with a simple rendition of what the Bogomol'nyi inequality [13] tells us about functionals which may be used in S-S' pair transport problems

$$V_E(\phi_F) - V_E(\phi_T) \equiv \Delta E_{gap} \propto L^{-1} \approx \tilde{\alpha} \tag{37}$$

for our fate of the false vacuum rendition of *S-S'* pair transport. A simple maximum-min argument allows us to approximate the false vacuum by :

$$\frac{\partial V}{\partial \phi} = 0,$$

$$\Rightarrow \phi_F \approx \left[\frac{\varepsilon^+}{\varepsilon^+ + 1}\right] \approx \varepsilon^+ \tag{38}$$

that is then tied in with the Bogomol'nyi inequality formulation of

$$L_E \approx \left[L_E|_{\phi=\phi_C} + \frac{4}{4^2} \cdot \frac{D \cdot \omega_P^2}{3!} \cdot (\phi_0^2 - \phi_C^2)^2\right] + \frac{1}{2} \cdot (\phi_0 - \phi_C)^2 \cdot \{\ \} \tag{39}$$

$$L_E \geq |Q| + \frac{1}{2} \cdot (\phi_0 - \phi_C)^2 \cdot \{\ \} \tag{40}$$

where

$$|Q| \to 0 \tag{41}$$

Due to a *topological* current argument (S-S' pairs usually being of opposite charge) and

$$\{\ \} \equiv \{\ \}_A - \{\ \}_B \equiv 2 \cdot \Delta E_{gap} \tag{42}$$

where if we pick :

$$\frac{(\{\ \} \equiv \{\ \}_A - \{\ \}_B)}{2} \equiv \Delta E_{gap} \equiv V_E(\phi_F) - V_E(\phi_T) \tag{43}$$

leads to a difference in energy levels between wells as given in the upper right hand side of Figure 1, where we made the link between the fate of the false vacuum hypothesis and the Bogomil'nyi inequality by requiring an equality in Eq. (43) above. Then we construct a gaussian wave functional which is congruent with respect to the false vacuum hypothesis[1] of Sidney Coleman. Here, we are looking at a quasi one dimensional expansion of $L_E$, which is done in a way that permits us to use a wick rotation, where we set

$$\alpha_2 \equiv \Delta E_{gap} \approx L^{-1} \quad (44)$$

and

$$\phi_T \geq \phi_0 \quad (45)$$

where

$$L_E \geq \frac{1}{2} \cdot (\phi_0 - \phi_C)^2 \cdot \{\ \} \quad (46)$$

$$\Psi_2 = c_2 \cdot \exp\left(-\alpha_2 \cdot \int d\tilde{x} [\phi_0 - \langle \phi \rangle_1]^2 \right)$$
$$\cong c_2 \cdot \exp\left(-\alpha_2 \cdot \int d\tilde{x} [\phi_T]^2 \right) \cong \Psi_{final} \quad (47)$$

$\phi_F \equiv <\phi>_1 \cong$ *very small value*, and in CDW we will be able to write $\phi_T \cong \phi_{2\pi} \equiv 2 \cdot \pi + \varepsilon^+$. But in general we will be usually be able to state that

$$|\phi_T| \geq |\phi_0| \quad (48)$$

Here for a future CDW example, we will have $\phi_0 \cong 2 \cdot \pi$ In any case though, we have that Eq. (41) and Eq. (47) together definitely fits the false vacuum hypothesis, and if we look at the CDW example where

$$\Psi_1 = c_1 \cdot \exp\left(-\alpha_2 \cdot \int dx [\phi_0 - \phi_{2\pi}]^2\right)$$

$$\cong c_1 \cdot \exp\left(-\alpha_1 \cdot \int dx [\phi_F]^2\right) \equiv \Psi_{initial} \tag{49}$$

We will assume this 2$^{nd}$ wave functional has $\alpha_2 \cong \alpha_1$. We will show the utility of this formalism in a first ever derivation of a current vs. electric field graph of charge density wave data in a follow up publication.

## VIII CONCLUSION

We should first note that we assume an almost instantaneous jump in the S-S' nucleation which allows us to reduce, by a dimension, a least action integral which permits easier calculation of the problem we were considering w.r.t. gaussian wave functionals. Our mathematical procedure is new and it was very helpful in simplifying how to present our wave functional results which tie in the fate of the false vacuum hypothesis of Sidney Coleman with construction of a gaussian wavefunctional presentation of S-S' pairs for the first time. This we will show show in a subsequent publication is important when we derive graphics which matched data sets for charge density waves. In addition we also present a description of entanglement in terms of the change in width of a S-S' pair which we believe is important and also is in its own way linked to how a topological charge argument may be used to isolate the gaussian contribution to a wave functional in terms of S-S' nucleation due to the Bogomil'nyi inequality [13]. We believe that the issues will yield important experimental insight and deserve to be investigated properly. We will do this in a subsequent application of the formalism to charge density wave transport problems.

How we construct $|T_{IF}|$ is a far-reaching problem, which could impact many areas of physics. For example, topological defects, such as flux vortices, play an important role in the cuprates and other type-II superconductors. Magnetic relaxation rates that depend weakly on temperature up to 20 K [32], or even decrease with temperature[8], suggest that Abrikosov vortices may tunnel over a wide temperature range. Moreover, the consistently low $I_c R_n$ products of cuprate Josephson devices suggest that Josephson vortex-antivortex pair creation may occur when the current is much smaller than the "classical" critical current $I_0 \sim \Delta/R_N e$.. Finally, the extraordinary rapidity of first-order phase transitions, such as the palpably visible nucleation of ice in supercooled water, suggest a possible similarity to the decay of the false vacuum.

# APPENDIX I: S-S' PAIRS IN TERMS OF S-S' WIDTH AND THE BOGOMOL'NYI INEQUALITY

We have used the Bogomol'nyi inequality[13] to isolate the Gaussian contribution to a wave functional treatment of S-S' pairs. We should note that the difference between $(\phi - \phi_0)^4$ and $(\phi^2 - \phi_0^2)^2$ comes out to be in this case $-4 \cdot \phi \cdot \phi_0 \cdot (\phi - \phi_0)^2$ put into the potential via the following manipulations:

$$V_E \cong C_1 \cdot (\phi - \phi_0)^2 - 4 \cdot C_2 \cdot \phi \cdot \phi_0 \cdot (\phi - \phi_0)^2 + C_2 \cdot (\phi^2 - \phi_0^2)^2 \tag{1}$$

So that then we can approximate our change in potential roughly as:

$$\Delta V_{Ci,f} \cong -4 \cdot \phi_0 \cdot \phi_{Ci,f} \cdot C_2 \cdot (\phi_{Ci,f} - \phi_0)^2 \tag{2}$$

s.t.

$$V_E\big|_{Ci,Cf} \cong (C_1 + \Delta V_{Ci,f}) \cdot (\phi_{Ci,f} - \phi_0)^2 + C_2 \cdot (\phi_{C,if}^2 - \phi_0^2)^2 \tag{3}$$

$$I_{1C_{i,f}} \equiv \int dx \cdot \alpha \cdot (\phi_{Ci,f} - \phi_0)^2 \tag{4}$$

vs.

$$I_{2Ci,f} \equiv \int dx \cdot \tilde{\alpha} \cdot (\phi_{Ci,f} - \phi_0)^2 \tag{5}$$

in a wave functional with an altered

$$\tilde{\alpha} \equiv \alpha - |\Delta V_{Ci,f}| \tag{6}$$

Comparing the behavior of both $I_1$ and $I_2$ gives an idea of the degree of overlap present in this system. That[30] we can say there is a maintained S-S' pair structure approximately in a *gaussian* state with only a shift in the relative width of the *gaussian* as represented by a shift in $\alpha$ and $\tilde{\alpha}$ values is a good argument in favor of a high degree of entanglement in this physical system, with direct implications as to the way we model current density in this system. A high degree of entanglement permits us to adopt Shih's analysis [30,31] of our maintained S-S' pair structure as acting as a *single object* that we argue is evidence we can use current density as proportional to the modulus of a tunneling Hamiltonian matrix element instead of the customary convention of current density as proportional to the square of the modulus of that tunneling Hamiltonian matrix element. In addition it also ties in directly with constructing the Gaussian wavefunctional and making it congruent w.r.t. the fate of a false vacuum hypothesis [1] as we discussed in Section V.

**FIGURE CAPTIONS :**

**Fig 1** Evolution from an initial state $\Psi_i[\phi]$ to a final state $\Psi_f[\phi]$ for a double-well potential (inset) in a 1-D model, showing a kink-antikink pair bounding the nucleated bubble of true vacuum. The shading illustrates quantum fluctuations about the initial and final optimum configurations of the field, while $\phi_0(x)$ represents an intermediate field configuration inside the tunnel barrier. The upper right hand side of this figure is how the fate of the false vacuum hypothesis gives a difference in energy between false and true potential vacuum values which we tie in with the results of the Bogomil'nyi inequality.

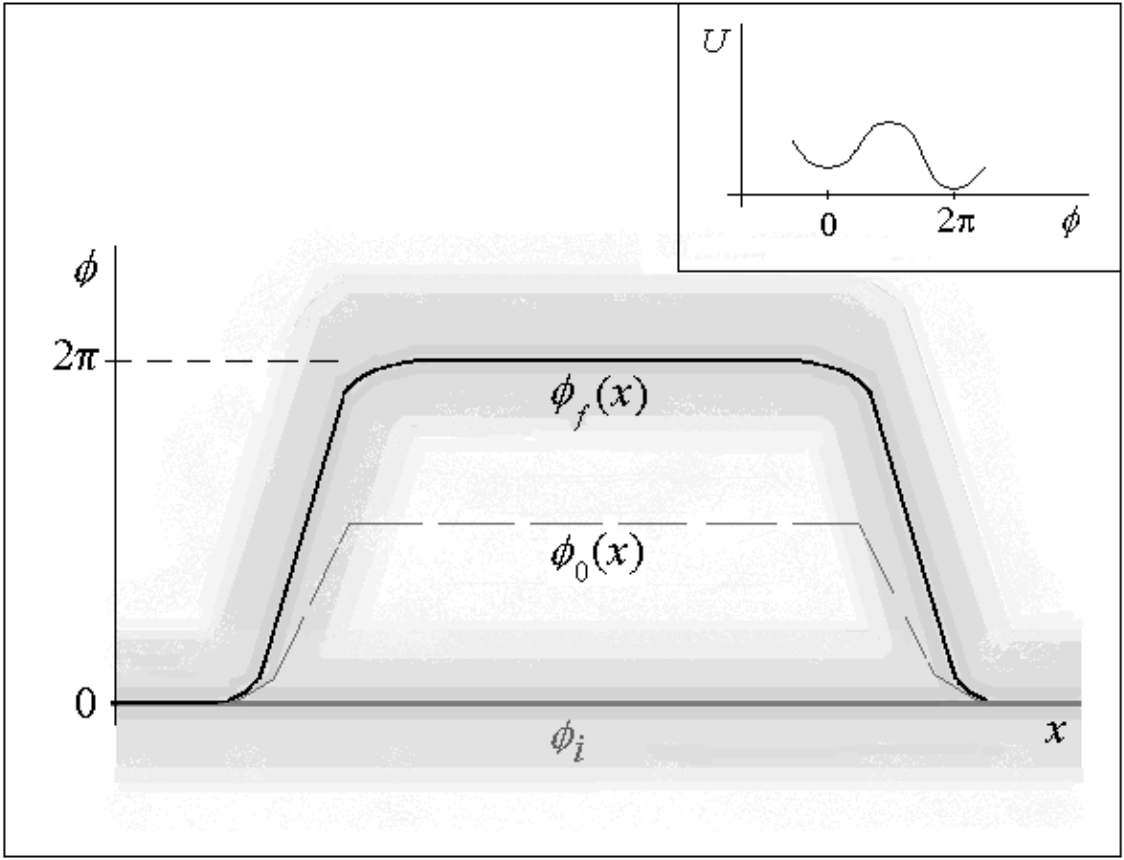

FIGURE 1

BECKWITH